\documentclass[a4paper,twoside,twocolumn]{article}
\pdfoutput=1

\newif\iftr 
\trtrue     

\usepackage{ifpdf}  
\ifpdf
    \usepackage[pdftex]{graphicx}
    \usepackage[colorlinks,linkcolor=blue,citecolor=blue,urlcolor=blue]{hyperref}
    \DeclareGraphicsExtensions{.pdf}
\else
    \usepackage{graphicx}
    \usepackage[hypertex]{hyperref}  
\fi

\usepackage{todonotes} 

\usepackage{
url       %
,fancyhdr 
,lastpage 
,enumerate 
,amsmath  
,amsfonts 
,amssymb  
, enumitem 
}
\usepackage[hang]{footmisc} 
\usepackage{balance}

\usepackage[noindent, arraystretch, fullpage]{setlengths} 
\usepackage{
own         
}

\usepackage{amsthm}

\newtheorem*{test*}{Test}
\usepackage{xfrac}

\graphicspath{{images/}}

\newcommand*{\metaauthori}{Bob Briscoe}
\newcommand*{\metaauthorii}{Olga Albisser}
\newcommand*{\metashorttitle}{AIMD-Friendliness}
\newcommand*{\metatitle}{Friendliness between AIMD Algorithms}
\newcommand*{\metano}{TR-BB-2021-002}
\newcommand*{\metakeywords}{Data Communication, Networks, Internet, Control, Congestion Control, TCP, Quality of Service, Rate Equality, Steady State, Fairness, Performance, Algorithm, Standards}

\newcommand*{\metamaili}{\href{mailto:research@bobbriscoe.net}{research@bobbriscoe.net}}
\newcommand*{\metamailii}{\href{mailto:olga@albisser.org}{olga@albisser.org}}
\newcommand*{\metaaddress}{}

\newcommand*{\metaversion}{01}
\newcommand*{\metadate}{17 May 2023}

\hypersetup{                       
     pdfauthor = {\metaauthori and \metaauthorii
     },
     pdftitle = {\metashorttitle},
     pdfsubject = {},
     pdfkeywords = {\metakeywords}
}%

\title{\metatitle}%
\author{\metaauthori%
\thanks{\metamaili, %
\metaaddress}%
\ %
\and \metaauthorii%
\thanks{\metamailii}%
}
\date{\metadate}%

\fancypagestyle{first}{%
\fancyhead[LO,RE]{}%
\fancyhead[LE,RO]{}%
\fancyhead[C]{}%
}%

\begin{document}
\bibliographystyle{alpha}%


\maketitle%
\thispagestyle{first}

\begin{abstract}
{\small\noindent%

This paper aims to provide a robust grounding for the additive increase factor used in the `TCP-Friendly' mode of the CUBIC congestion control algorithm.}      
\end{abstract}
\section{Introduction}\label{Introduction}

The first IETF RFC to define the CUBIC~\cite{Rhee18:Cubic_RFC} congestion control algorithm (CCA) was based on the original paper introducing CUBIC~\cite{Ha08:cubic}. For `TCP-friendly' mode, both draw on an equation in an ACIRI technical report~\cite{Floyd00:Eqn_v_AIMD_cc} when they specify the additive increase factor. The derivation of the equation in that technical report has been called into question. So this report aims to more rigorously derive and evaluate the required additive increase factor. The ACIRI report assumes a deterministic dropping (or ECN-marking) algorithm at the bottleneck, which limits its applicability. Also it attempts to validate the theoretical formula empirically by simulation using a RED gateway at the bottleneck, but it results in flow rates that are inexplicably different (by a factor of more than 2\(\times\)) when they should be the same.

Below, an equation for the additive increase factor is derived without the assumption of deterministic dropping. Instead it is assumed that drops are synchronized between flows, which is typically the case for tail-drop queues. The resulting equation turns out to be the same as that in the technical report~\cite{Floyd00:Eqn_v_AIMD_cc}. However, the derivation here is different. It has a straightforward geometric interpretation and it relies on fewer assumptions and no approximations; it considers variation of the RTT explicitly and it does not use loss probability at all.

The present paper is not intended to be ambitious or insightful, just pedestrian and rigorous. \cite{Bansal01:Binom_cc} is recommended for insight into the wider set of `TCP-friendly' algorithms, but it does not go into depth on the simple linear cases analysed here.

\section{Terminology}\label{Terminology}

Nowadays, CUBIC's TCP-friendly mode is more accurately known as Reno-friendly mode, given its flow rate is intended to match that of the Reno CCA, and given that it is irrelevant which wire protocol is used, whether TCP, QUIC, SCTP, etc. The term C-Reno will be used for CUBIC in Reno-friendly mode.

This paper uses the variables defined below:
\begin{description}[nosep]
	\item[\(a\)]: Additive increase factor;
	\item[\(b\)]: Multiplicative decrease factor;
	\item[\(j\)]: Round index;
	\item[\(J\)]: Rounds per sawtooth cycle;
	\item[\(R(j)\)]: Round trip time (RTT);
	\item[\(W(j)\)]: Congestion window;
	\item[\(\widehat{W}\)]: Maximum \(W\);
	\item[\(r(j)\)]: Packet rate;
	\item[\(X_r\)]: Reno variant of any variable \(X\);
	\item[\(X_c\)]: C-Reno variant of any variable \(X\).
\end{description}

\section{AIMD-Friendliness}\label{AIMD-Friendliness}

Consider two types of Additive Increase Multiplicative Decrease (AIMD) flow with parameters (\(a_r, b_r\)) and (\(a_c, b_c\)) competing at a bottleneck, under the following assumptions:
\begin{itemize}[nosep]
	\item The buffer is large enough not to drain completely, even if all flows reduce simultaneously (this assumption is relaxed later).
	
	\item All other factors of all the flows, particularly packet size and base RTT, are equal. When flows sharing the same bottleneck queue all have the same base RTT, they all have the same RTT, \(R(j)\),  at every stage, \(j\), of their sawtooth cycles. 	
	
	\item The bandwidth-delay product (BDP) is low enough for all flows to remain in their AIMD mode throughout the cycle.

	\item All the flows have run long enough to converge to a steady state.

	\item All flows only respond to the presence of loss, not its extent.
\end{itemize}

\begin{figure*}
	\centering
	\includegraphics[height=6.5cm]{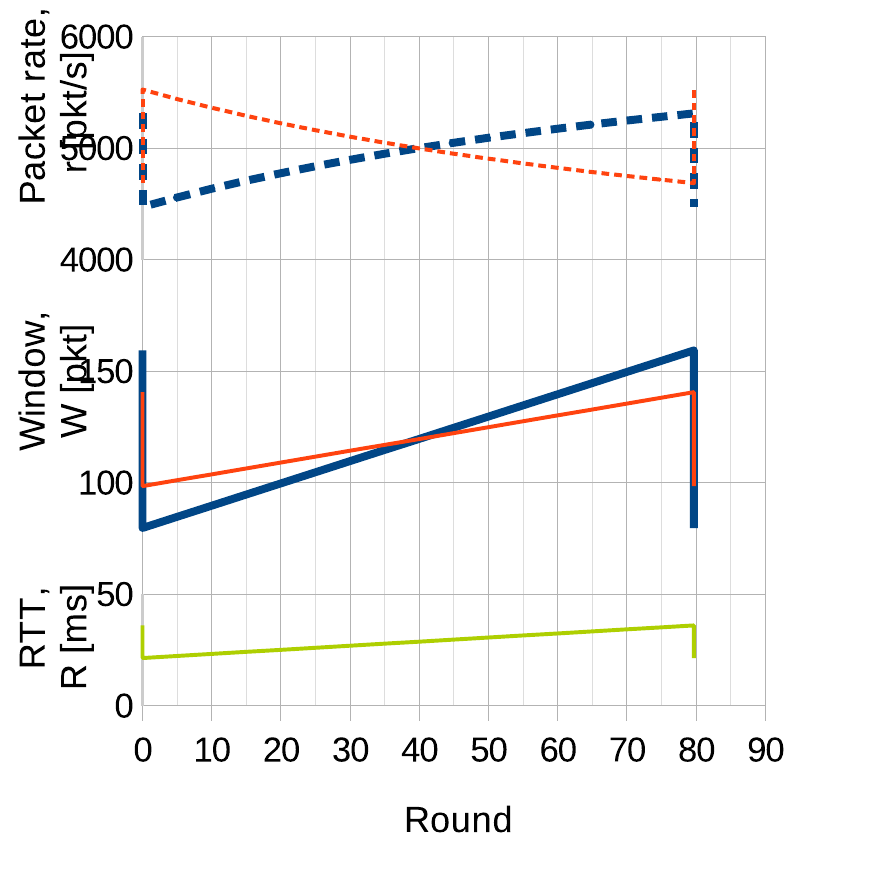}
	\includegraphics[height=6.5cm]{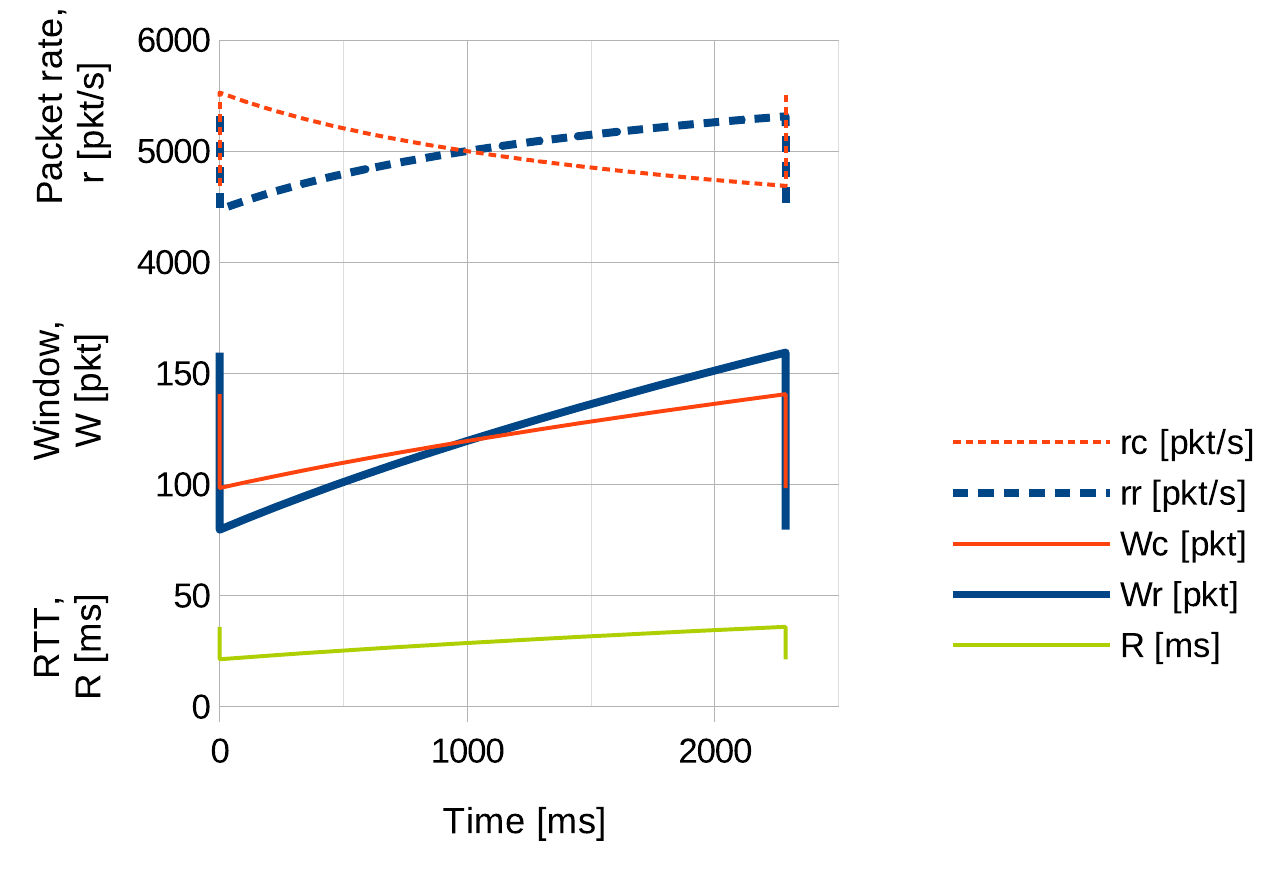}
	\caption{One synchronized sawtooth cycle of C-Reno and Reno plotted wrt.\ round trips (left) and wrt.\ time (right)}\label{fig:creno-synch}
\end{figure*}

\subsection{Synchronized (Tail-Drop) Case}\label{Synch}

For this case, it is additionally assumed that:
\begin{itemize}[nosep]
	\item All the flows are synchronized so that, whenever one flow experiences loss the others do too;

	\item All flows experience at least one loss at each congestion event (relaxation of this assumption is discussed later);
\end{itemize}

\paragraph{Steady state:} For each flow, the additive increase of a cycle balances with its multiplicative decrease from the max, \(\check{W}/b\), to the min, \(\check{W}\). 
\begin{align}
	a_r J &= \widehat{W}_r - b_r\widehat{W}_r\notag\\
	      &= \widehat{W}_r(1-b_r)\label{eqn:ssr}\\
	a_c J &= \widehat{W}_c(1-b_c)\label{eqn:ssc}
\end{align}

\paragraph{Flow rate equality:} Given the parameters \(a_r, b_r, b_c\) the aim is to derive \(a_c\) such that each flow's average rate is the same. This is equivalent to each flow transferring the same number of packets over a cycle. 

As a cycle progresses, the RTT grows. So to derive the number of packets transferred over a cycle, the packet rate has to be weighted by the RTT in each cycle before being summed:
\begin{align}
	\sum_{j=0}^{J-1} r_c(j) R(j) &= \sum_{j=0}^{J-1} r_r(j) R(j)\notag\\
	\sum_{j=0}^{J-1} W_c(j) &= \sum_{j=0}^{J-1} W_r(j)\label{eqn:cwnd_equality}\\
	\sum_{j=0}^{J-1} b_c\widehat{W}_c + a_c j &= \sum_{j=0}^{J-1} b_r\widehat{W}_r + a_r j\notag\\
	Jb_c\widehat{W}_c +\frac{J^2 a_c}{2} &= Jb_r\widehat{W}_r +\frac{J^2 a_r}{2}\notag\\
\intertext{Dividing through by \(J\) and substituting from \autoref{eqn:ssr} \& \autoref{eqn:ssc}:}
    \widehat{W}_c \left(b_c + \frac{(1-b_c)}{2}\right) &= \widehat{W}_r \left(b_r + \frac{(1-b_r)}{2}\right)\notag\\
	\frac{\widehat{W}_c}{\widehat{W}_r} &= \frac{(1+b_r)}{(1+b_c)}\label{eqn:fre}
\end{align}
Returning to the steady state equations, we can divide \autoref{eqn:ssc} by \autoref{eqn:ssr}, then substitute from \autoref{eqn:fre}:
\begin{align}
	\frac{a_c}{a_r} &= 
	                 \frac{\widehat{W}_c}{\widehat{W}_r}\frac{(1-b_c)}{(1-b_r)}\notag\\
	                &= \frac{(1-b_c)}{(1+b_c)}\frac{(1+b_r)}{(1-b_r)}\label{eqn:friendly}
\end{align}
Plugging in Reno's AIMD factors, \(a_r=1, b_r=1/2\):
\begin{equation}
                \boxed{a_c = \frac{3(1-b_c)}{(1+b_c)}}\label{eqn:reno-friendly}
\end{equation}
\begin{figure*}
	\centering
	\includegraphics[height=6.5cm]{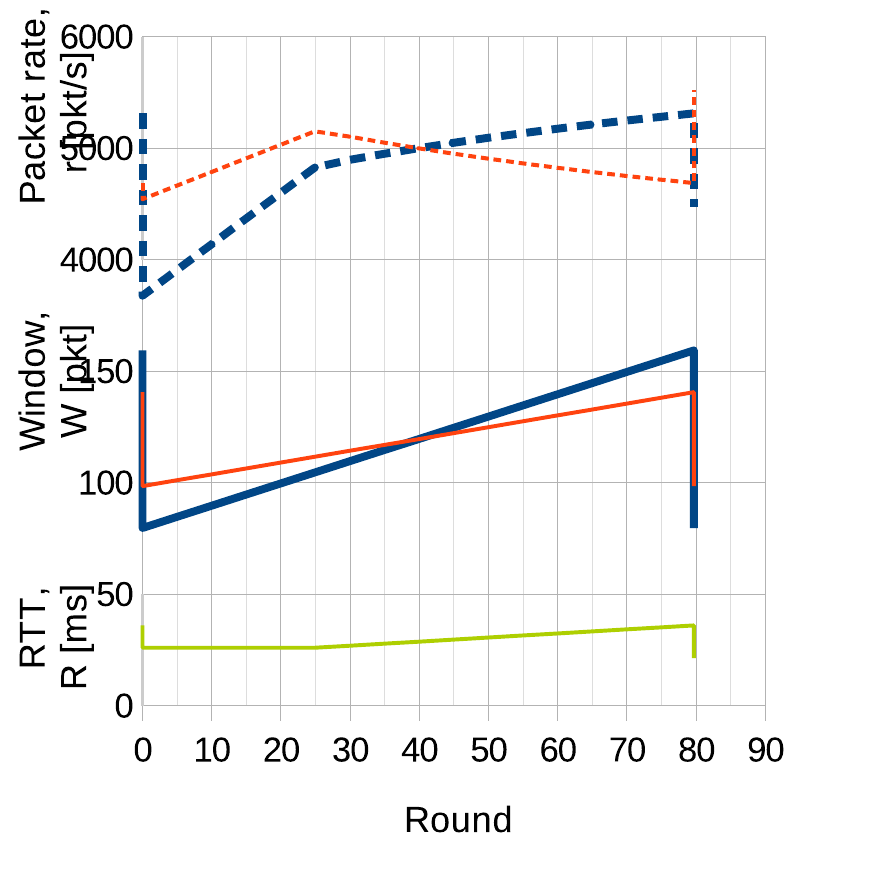}
	\includegraphics[height=6.5cm]{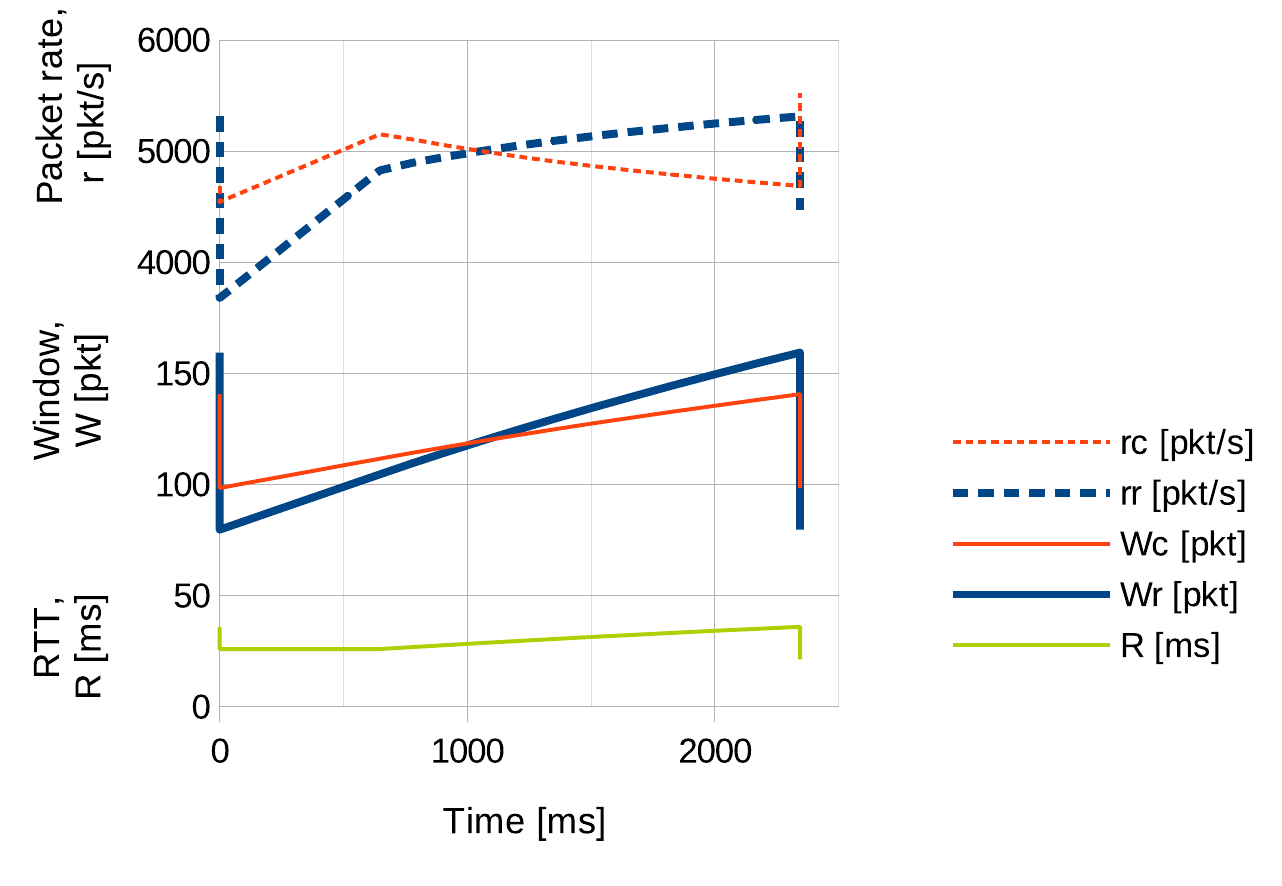}
	\caption{Similar synchronized sawtooth cycle of C-Reno and Reno to \autoref{fig:creno-synch}, but with a 11\,ms buffer that is too shallow to accommodate both sawteeth at once.}\label{fig:creno-synch-shallow}
\end{figure*}And plugging in the multiplicative decrease factor of C-Reno recommended in \cite{Rhee18:Cubic_RFC}, \(b_c=0.7\):
\begin{align*}
               a_c  &= 9/17\\
                    &\approx 0.53.
\end{align*}

\subsubsection{Geometric interpretation} 
\autoref{fig:creno-synch} shows one flow each of C-Reno and Reno competing over one synchronized sawtooth cycle. Superficially, the whole derivation of \(a_c\) above can be derived from simple triangle geometry, by drawing congestion window sawteeth that increase linearly wrt.\ round trips (mid-left) then setting the mid-points of the two ramps to the same height. Then \autoref{eqn:fre} gives the ratio between the heights of the bases of the triangles, and \autoref{eqn:friendly} gives the ratio of the heights of the triangles themselves.

However, it is not enough to merely assert that the average heights of these sawteeth are equal, as \cite{Floyd00:Eqn_v_AIMD_cc} does.%
\footnote{It is unlikely that the additional initial steps given here were implicit but unstated, because a high-level averaging approach was used.}
Strictly, it is necessary to start from the goal of equal flow rates averaged over time, as the above analysis does. Given RTT grows throughout the cycle, the plots of flow-rate against time (top right in \autoref{fig:creno-synch}) stretch out more to the right, forming concave curves. It is not at all obvious how to equate the averages of these two curves until they are weighted by round trip duration, which transforms them into the linear plots of window wrt.\ rounds (mid-left), as in \autoref{eqn:cwnd_equality}.

It is also interesting to note from \autoref{fig:creno-synch} that C-Reno's packet rate \emph{decreases} as its window increases over the sawtooth. This is because the competing Reno flow causes the RTT to grow faster than would be the case with only C-Reno flows. 

If the buffer is not deep enough to hold all the synchronized sawteeth, it will be empty during the early part of the sawteeth. Then both flows will under-perform during the early part of the cycle when C-Reno would have achieved its highest packet rate and Reno would have achieved its lowest, as shown in \autoref{fig:creno-synch-shallow}. Nonetheless, the rate of both flows reduces proportionately, so the ratio between their flow rates remains unaltered.

\subsubsection{Synchronized losses?}
We will now question the assumption that each flow always catches at least one loss at each congestion event. We still consider two flows with equal base RTTs: 1 Reno and 1 C-Reno. Between responses to losses, the queue grows inexorably by \((a_r + a_c \approx 1.53)\) pkt/RTT. Every time the buffer fills, one packet has to be dropped, but the queue continues to grow by 1.53 segments during the next round trip (until the resulting response reaches the queue). So it is likely that another packet will have to be discarded within the same RTT as the first.

The likelihood that a particular flow catches any one of the losses depends on its packet rate relative to the other.\footnote{Irrespective of how rapidly the flow's own window grows.} That is,
\begin{align}
	\frac{p_r}{p_c} = \frac{\widehat{r_r}}{\widehat{r_c}}.\label{eqn:p_ratio1}
\end{align}

If the flows both have the same average window (the goal in \autoref{fig:creno-synch})
then, by \autoref{eqn:fre} (or triangle geometry), the ratio between the packet rates of the two flows when they both reach their max is
\begin{align}
\frac{\widehat{r_r}}{\widehat{r_c}} &= \frac{(1 + b_c)}{(1 + b_r)}\notag\\
        &= \frac{1.7}{1.5} &&\approx 1.13.\label{eqn:r_ratio}
\end{align}

When a loss occurs, from \autoref{eqn:p_ratio1} \& \autoref{eqn:r_ratio}, we can say that:
\begin{align}
	p_r / p_c = 17/15\label{eqn:p_ratio}\\
\mathrm{and}\qquad p_r + p_c = 1,
\end{align}
because one or the other flow was hit with certainty.

Therefore
\begin{align*}
p_r &= 17/32 &\approx 53\%\\
p_c &= 15/32 &\approx 47\%
\end{align*}
Then, for example, if there are two losses during a congestion event, the probabilities of each combination of two losses are:
\begin{align*}
p_{rr}              &= (17/32)^2               &&\approx 28\%\\
p_{rc} \lor p_{cr}  &= 17*15/32^2 + 15*17/32^2 &&\approx 50\%\\
p_{cc}              &= (15/32)^2               &&\approx 22\%,
\end{align*}
where \(p_{ij}\) is the probability of a loss from flow \(i\) then \(j\).

When there are two losses in a round and the same flow is hit twice, it doesn't reduce any more than if it's hit once, but the other flow doesn't reduce at all. So C-Reno is somewhat more likely than Reno to not get hit in some round. In such cases, only the Reno flow would reduce, then the queue would continue to grow by \((a_r + a_c \approx 1.53)\) pkt/RTT, so the next cycle would be shorter and the C-Reno flow would be much more likely to be hit when it next filled the buffer --- and more likely to be hit twice.

It would be possible to calculate the average rate of each type of flow by calculating the probabilities of each chain of events programmatically. However, such precision is unnecessary. For the case of tail-drop buffers, it will be sufficient to say:
\begin{itemize}
	\item either that the AI factor of C-Reno should be slightly lower than that derived from \autoref{eqn:reno-friendly} to make C-Reno and Reno flow rates more precisely equal;
	\item or that the average rate of C-Reno flows will be slightly higher in comparison with Reno flows, if \autoref{eqn:reno-friendly} is used.
\end{itemize}
``Slightly''  means within 10\%, which conservatively accommodates the difference between \(p_{rr}\) and \(p_{cc}\).

\subsubsection{Empirical Results: Tail Drop}\label{empirical-pfifo}

\begin{figure}
	\centering
	\includegraphics[width=\linewidth]{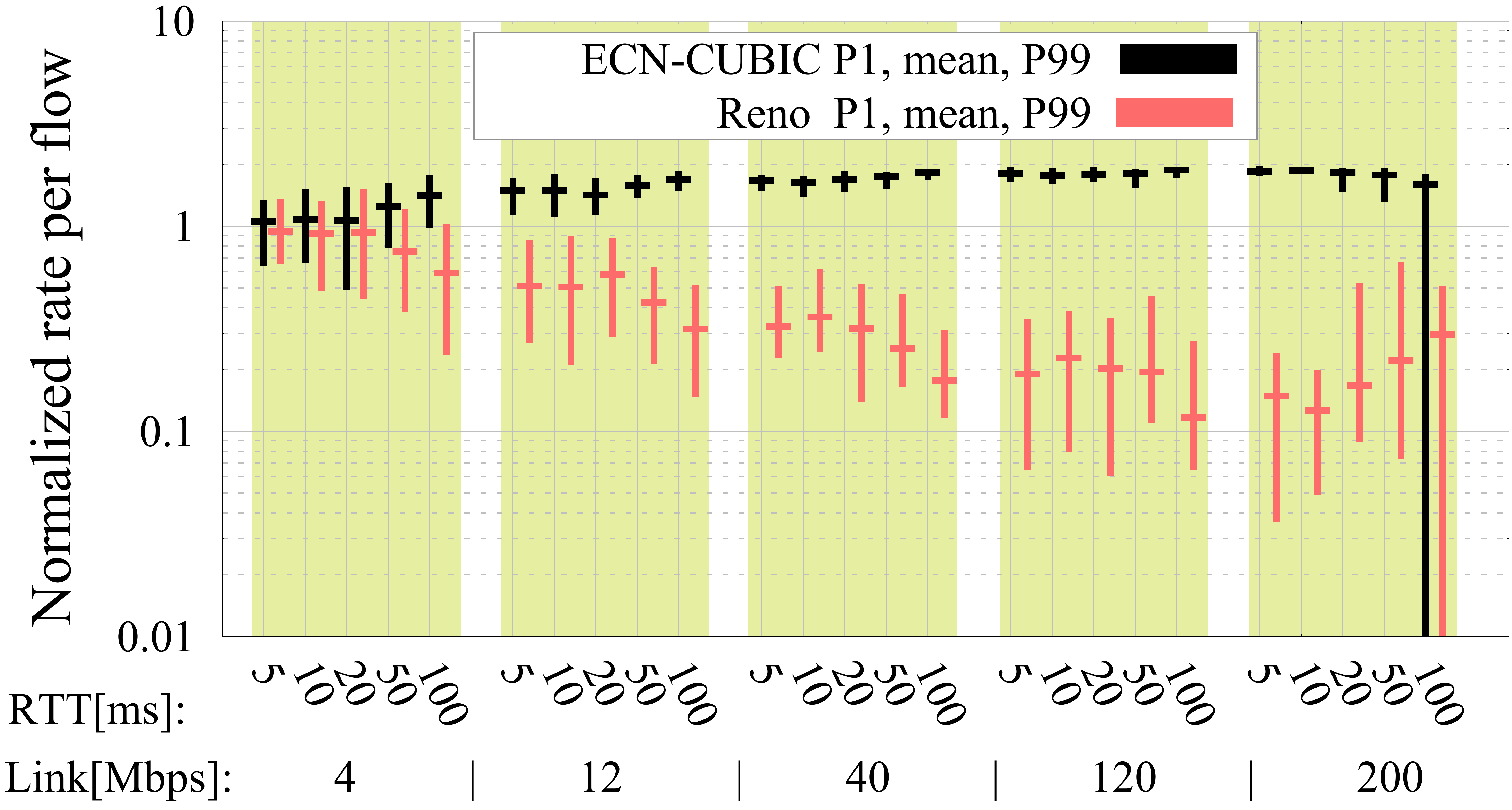}
	\caption{Empirical results comparing the flow rates of CUBIC and Reno competing 1:1 over a tail-drop queue in a 200\,ms (`bloated') buffer with MD factor \(b_c=0.7\) and AI factor \(a_c=0.53\) for CUBIC's Reno-friendly algorithm}\label{fig:ts_1-1_erpfifoRmax200ms}
\end{figure}

\paragraph{Testbed set-up:} A single long-running C-Reno flow with MD factor \(b_c = 0.7\) and AI factor \(a_c=1.53\) (the defaults in Linux derived using \autoref{eqn:reno-friendly}) was tested against a single Reno flow over a tail-drop buffer. The buffer was at the bottleneck of a range of 25 different Ethernet paths with five different link rates and five different RTTs, as shown on the \(x\)-axis of \autoref{fig:ts_1-1_erpfifoRmax200ms}.\footnote{In each run, flow rate measurement was delayed until the flows had converged, then taken every second for 250\,s. Percentiles were derived from the resulting 250 measurements. The two types of flow were sent from two sending hosts to two receiving hosts via a bridge and a bottleneck node. All machines were running Linux kernel v5.10.31. All CCAs were configured with default parameters.} 

Two buffer sizing approaches were evaluated: `bloated' and `tuned'. In both the buffer was sized dependent on the link rate, but in the `bloated' case it held  200\,ms and in the `tuned' case only 25\,ms.%
\footnote{The buffer was not sized to the RTT of each flow. Instead, to better match operational practice, it was sized to a worst-case RTT, given the RTT of each flow is hard to measure in the network. Then two different definitions of worst-case are compared.}
That is, buffer size \(B\:[\textrm{byte}] = C\widehat{R}/8\), where \(C\) [b/s] is the link rate serving the buffer and \(\widehat{R}\) is either 0.2\,s or 0.025\,s. 

The `bloated' case represents the outdated rule of thumb where an operator allows for 1\,BDP of buffer at the maximum feasible RTT, where 200\,ms approximates the RTT in glass fibre around half the earth's circumference and back. The `tuned' case is tailored to CUBIC's MD factor of 0.7 and to its slow-start behaviour, on the assumption that the base RTT of most Internet traffic is under 50\,ms.

In all cases the results are displayed as whisker plots that show 1\textsuperscript{st} \%-ile (P1), mean and 99\textsuperscript{th} \%-ile (P99) normalized flow rate. Normalized flow rate is defined as the flow rate relative to \(\sfrac{1}{N}\) of the capacity, where \(N\) is the total number of flows.

The shaded background of part of the plots indicates those cases where CUBIC is not in it's Reno-friendly mode. Darker shading represents true CUBIC mode, while lighter shading indicates where additive increase transitions from Reno-friendly to true CUBIC part-way up each saw-tooth. The switch-over RTT is taken from equation (6) in \cite{Briscoe21c:pi2param} and it is assumed that all sawteeth peak at the tail of the buffer.

\paragraph{Interpretation of Results}

In the `bloated' buffer case, \autoref{fig:ts_1-1_erpfifoRmax200ms} shows that none of the CUBIC flows are ever in Reno-friendly mode. This tells us nothing relevant to the present paper, but it shows that CUBIC can significantly outcompete Reno (up to about 15:1 rate ratio) when not in Reno-friendly mode,%
\footnote{These experiments ensured that both flows remained in steady state, but the argument for CUBIC rests on Reno being more sensitive to disturbances that knock it out of its steady state~\cite{HA2007:CUBIC_bg}.}
although at least Reno does not starve at low link rates.

\begin{figure}
	\centering
	\includegraphics[width=\linewidth]{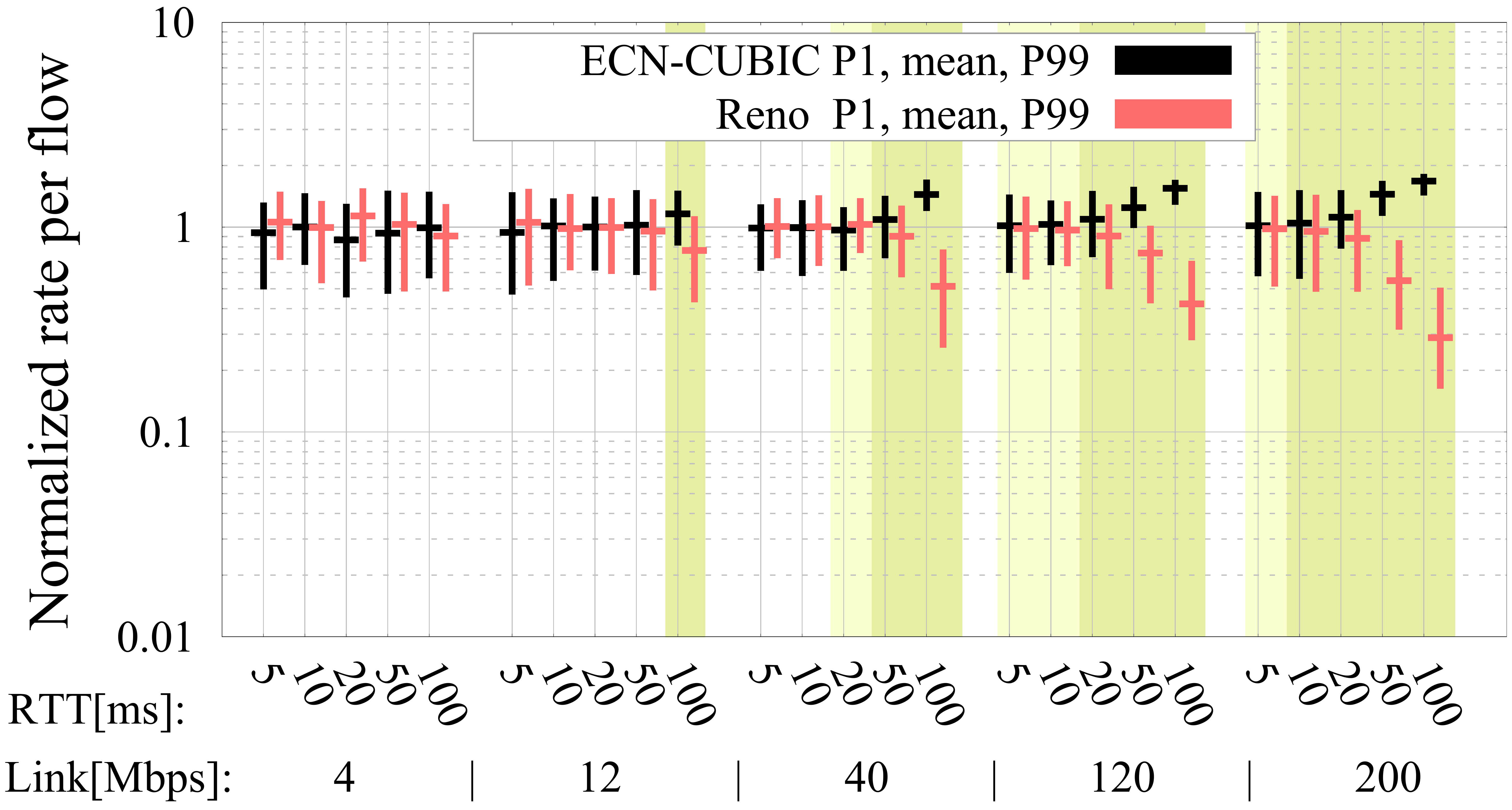}
	\caption{Empirical results with same set up as \autoref{fig:ts_1-1_erpfifoRmax200ms} except buffer tuned to 25\,ms}\label{fig:ts_1-1_erpfifoRmax25ms}
\end{figure}

In the `tuned case', \autoref{fig:ts_1-1_erpfifoRmax25ms} shows that CUBIC is either in or partly in its Reno-friendly mode over a larger part of the scenario space (respectively unshaded or lightly shaded background). In all such cases, the mean normalized flow rate of both flows is close to 1. This validates the model of tail drop behind \autoref{eqn:reno-friendly}, at least for the case when \(b_c=0.7\).

Outside its Reno-friendly mode, CUBIC increasingly dominates Reno, but that is outside the AIMD-specific focus of the present paper.

\subsection{Desynchronized (AQM) Case}\label{AQM}
\begin{figure*}
	\centering
	\includegraphics[height=6.4cm]{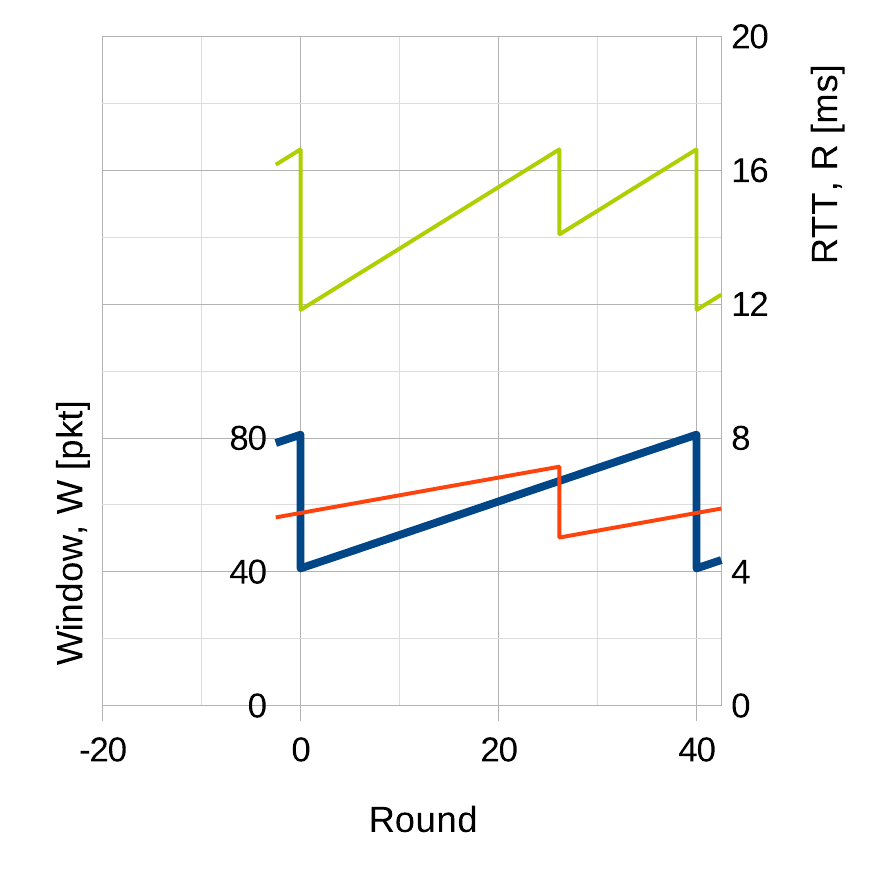}
	\includegraphics[height=6.4cm]{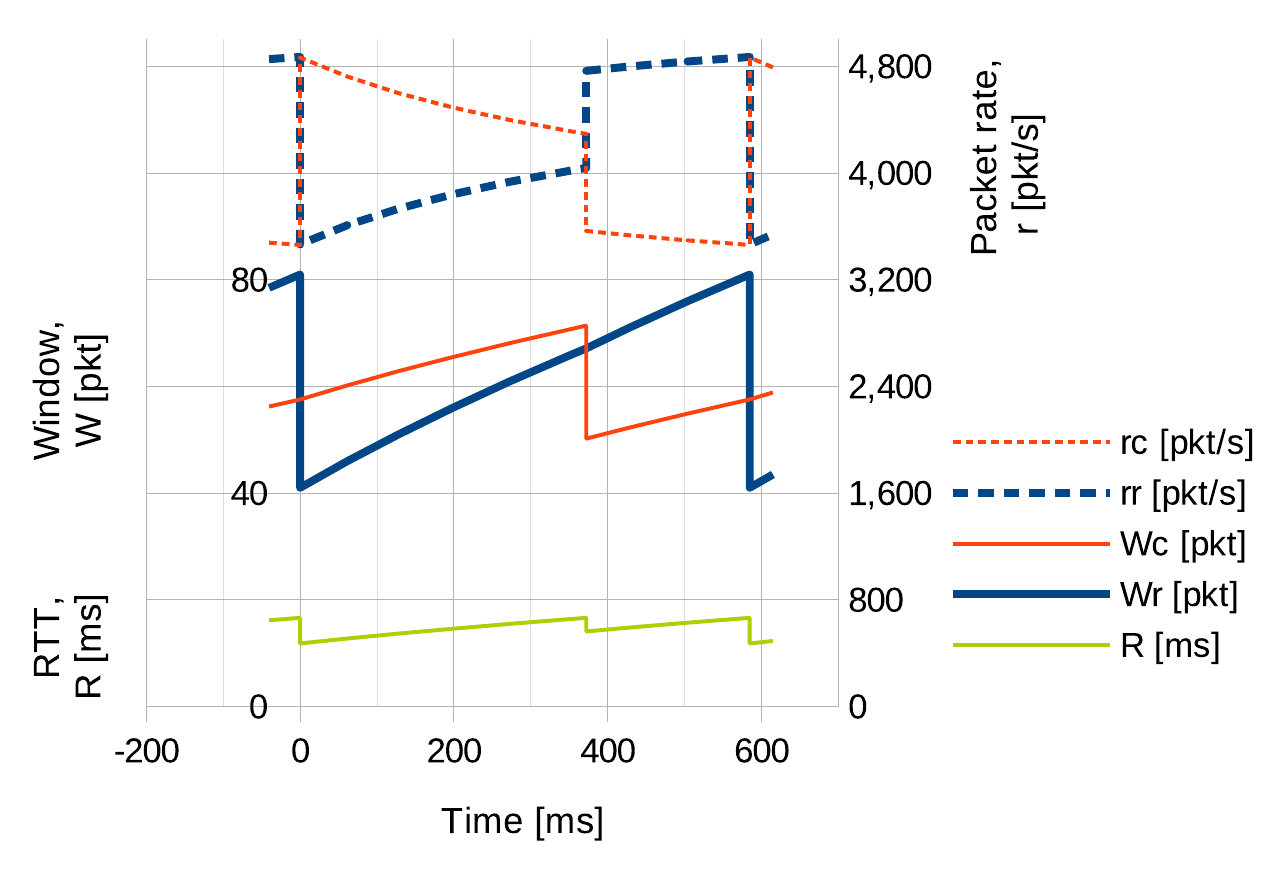}
	\caption{One desynchronized sawtooth cycle of C-Reno and Reno plotted wrt.\ round trips (left) and wrt.\ time (right)}\label{fig:creno-determ}
\end{figure*}

For this case, instead of the assumption of synchronization, it is assumed that:
\begin{itemize}[nosep]
	\item As the queue grows, Active Queue Management (AQM) at the bottleneck selects single packets to drop or mark, so that the congestion responses of each flow tend not to coincide.\footnote{In desynchronized cases, the RTT varies less than in synchronized cases---on average the amplitude is respectively \(1/\sqrt{n}\) vs.\ \(n\) times that of a single flow, where \(n\) is the number of flows~\cite{Appenzeller04:Sizing_buffers}. Therefore, the first assumption listed in \S\,\ref{AIMD-Friendliness} (that the buffer is large enough not to drain completely) is much more likely to hold in desynchronized cases.}
\end{itemize}

The AQM case is harder to analyse than the synchronized case with tail-drop. Superficially, one could use the transformation from equal average flow rates (wrt.\ time) into equal window (wrt.\ rounds). However, there is no guarantee that the number of rounds per cycle, \(J\) is the same in each case.

If we assumed it was, we would end up with \autoref{eqn:reno-friendly} for C-Reno's additive increase factor \(a_c\). Then, as shown in \autoref{fig:creno-determ}, the phasing between the sawteeth would evolve so that the queue reached roughly the same depth before each reduction, i.e.\ the tips of the RTT sawteeth will all align at roughly the same level---the operating point of the AQM. 

However, although \autoref{fig:creno-determ} shows the sawtooth reductions alternating Reno -- C-Reno -- Reno, this need not be the case. In the cycle after 400\,ms in the top-right plot, the ratio between C-Reno's and Reno's packet rates is about 51:49. So it is nearly as likely that the AQM will hit a Reno packet as a C-Reno packet, causing Reno to reduce twice in a row. If the AQM did hit C-Reno around 400\,ms, at around 600\,ms the ratio would be about 42:58, making Reno more likely to be hit. 

The probability of each outcome at each stage could be derived programmatically (a Markov chain), but it will suffice to test empirically whether the model in \autoref{eqn:reno-friendly} fits the AQM case. We can at least suppose that, if \(a_c\) is set as for the synchronized case (\autoref{eqn:reno-friendly}), an AQM is likely to hit Reno somewhat more often than C-Reno.

\subsubsection{Empirical Results: PIE AQM}

\paragraph{Testbed set-up:} To test the AQM case, similar experiments to those in \S\,\ref{empirical-pfifo} were run, but this time with a PIE AQM not a tail-drop buffer.\footnote{On a v5.10.31 Linux kernel with default parameters.}

\begin{figure}
	\centering
	\includegraphics[width=\linewidth]{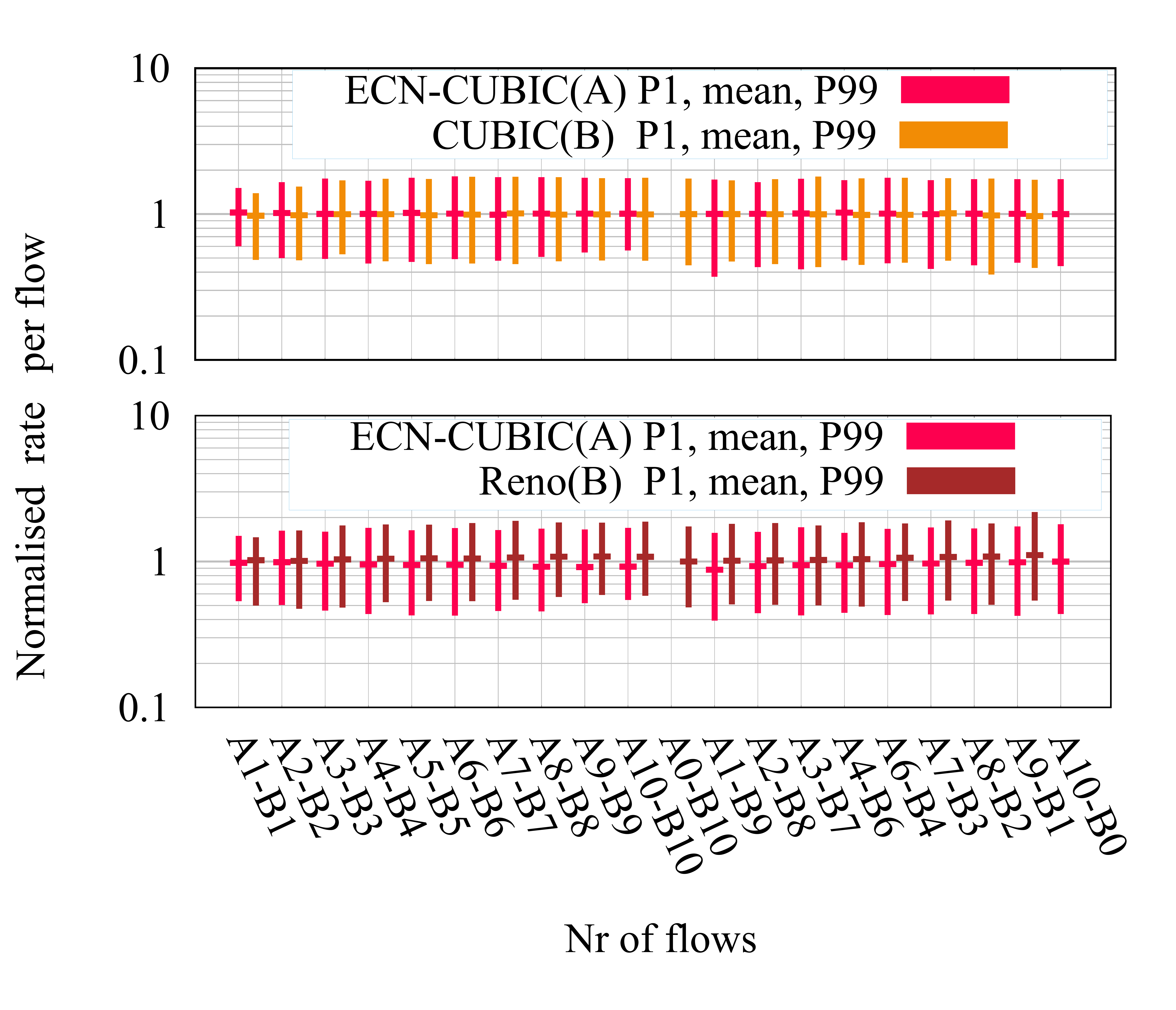}
	\caption{Average flow rates of different numbers of long-running CUBIC and Reno flows. AQM: PIE; link capacity: 40\,Mb/s; Base RTT: 10\,ms.}\label{fig:creno-v-reno-PIE}
\end{figure}

Also, various combinations of numbers of each flow type were tested, as shown along the \(x\)-axis of \autoref{fig:creno-v-reno-PIE}. For instance, in the bottom plot of \autoref{fig:creno-v-reno-PIE}, A2-B8 means 2 ECN-CUBIC flows vs. 8 (non-ECN) Reno flows. The top plot of the same figure shows the results of equivalent control experiments where the Reno flows were replaced by non-ECN CUBIC flows. This served to check that Reno and C-Reno flows were equally agressive against the same set of ECN C-Reno flows.

the ECN-capability of the C-Reno flows was not altering their aggressiveness.

All the different combinations of flow numbers were run over a path with base RTT 10\,ms over a 40\,Mb/s bottleneck, which was chosen to keep CUBIC in its Reno-friendly mode. 

See \S\,\ref{empirical-pfifo} for the meanings of the whiskers on the plots, the `normalized flow rate' metric and the shaded regions of the background.%
\footnote{With an AQM, it would seem less straightforward to determine whether CUBIC will have started to transition from C-Reno to pure CUBIC mode at a particular base RTT, because it is hard to estimate where the tips of the sawteeth sit in relation to the 15\,ms AQM target (see \S\,3.3 of \cite{Briscoe21c:pi2param} for background). However, at the selection of base RTTs shown, it turned out that using any feasible estimate didn't change whether CUBIC was in transition. In other words, the mode CUBIC is in is fairly insensitive to this estimate.}

\begin{figure}
	\centering
	\includegraphics[width=\linewidth]{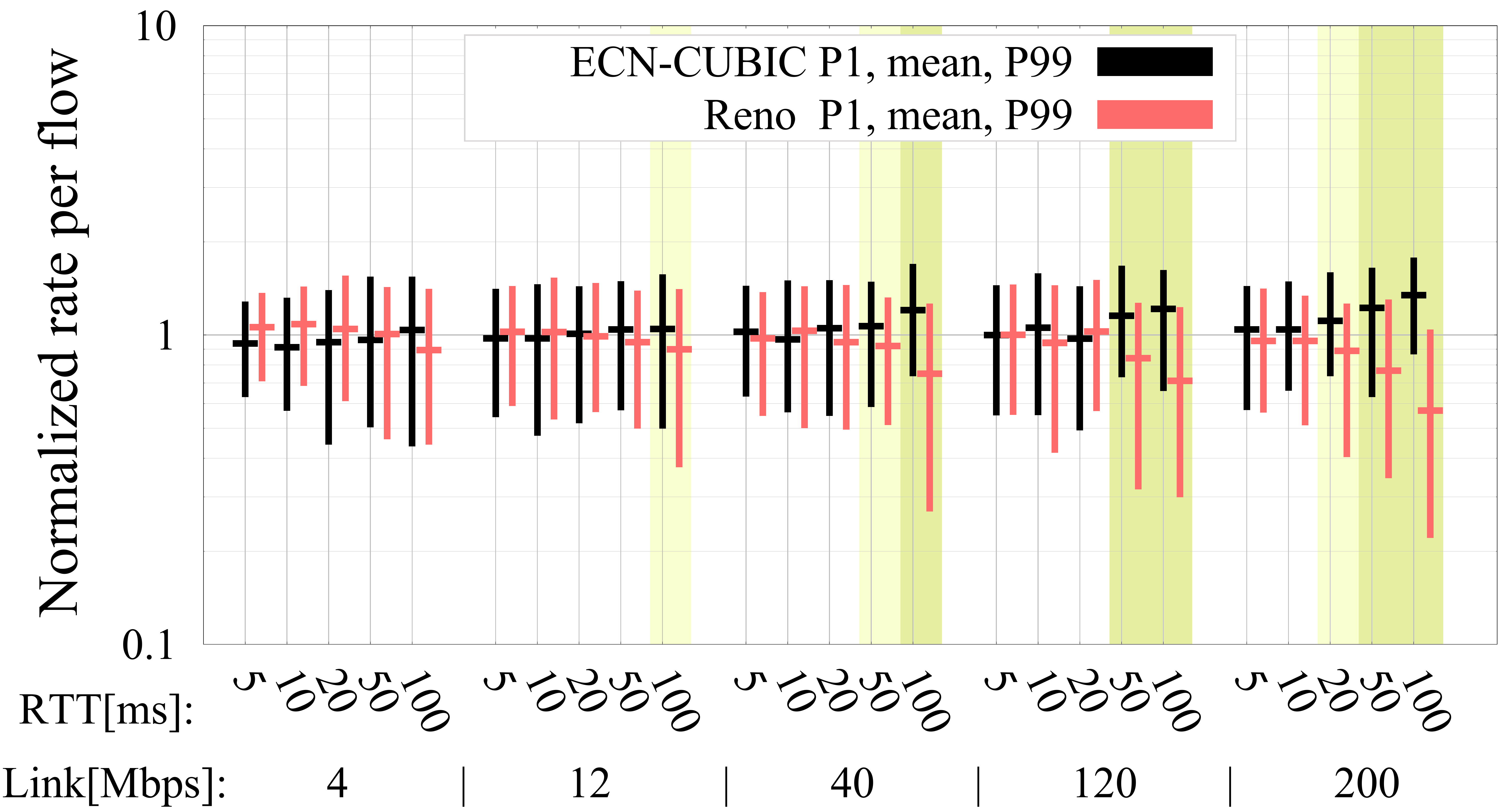}
	\caption{Empirical results with set up similar to \autoref{fig:ts_1-1_erpfifoRmax200ms} except over a PIE AQM}\label{fig:ts_1-1_erpie}
\end{figure}

\paragraph{Interpretation of Results:} In the 1 CUBIC vs.\ 1 Reno tests (\autoref{fig:ts_1-1_erpie}) over a PIE AQM, it can be seen that C-Reno and Reno share the bandwidth roughly equally. However, as BDP increases, CUBIC can be seen to start taking a greater share of capacity, as it increasingly operates beyond its Reno-friendly mode.\footnote{Further discussion of pure CUBIC mode is outside the scope of the present paper, which focuses on AIMD.}

Thus, in practice, the AI factor derived from the tail drop model keeps the flow rates roughly equal whether the bottleneck is tail drop or a PIE AQM. 

In the multi-flow tests over the PIE AQM (\autoref{fig:creno-v-reno-PIE}), it can be seen that C-Reno gives roughly the same rate as Reno over the full range of tests. Indeed, if anything, C-Reno seems to be slightly less aggressive than Reno. However, the discrepancy is hardly visible, and more than an order of magnitude smaller than the range of the ratio's natural variability between its P1 and P99. 

\section{Conclusion}\label{Conclusion}

This report provides:
\begin{itemize}[nosep]
	\item a formula (\autoref{eqn:reno-friendly}) for the additive increase parameter of an AIMD algorithm as a function of its chosen multiplicative decrease factor that should maintain an equal flow rate with another AIMD flow, specifically a Reno flow. 
	
	\item a derivation of the formula that relies on fewer assumptions and is more rigorous than that in Floyd \emph{et al}~\cite{Floyd00:Eqn_v_AIMD_cc}. It applies to tail drop buffers whereas that in Floyd \emph{et al} relied on an AQM with deterministic marking. Nonetheless the formula turns out to be the same.
	
	\item a testbed validation of the formula (at least for the case where the MD factor is 0.7) over a range of 25 different path characteristics (5 rates and 5 base RTTs) with a tail-drop bottleneck buffer.
	
	\item a testbed evaluation of the formula over the same range of scenarios but with a PIE AQM at the bottleneck. This shows that the AI factor works correctly over a PIE AQM, even though it was derived assuming tail-drop.
\end{itemize}

These result show that, with the widely deployed decrease factor of \(b_c=0.7\), the Reno-Friendly mode in CUBIC is sufficiently well modelled by \autoref{eqn:reno-friendly} that the Additive Increase factor it produces (\(a_c = 0.53\)) ensures that TCP CUBIC competes roughly equally with Reno across its intended operating range, whether with a tail-drop queue or a single-queue AQM (PIE) at the bottleneck.      

\newpage
\addcontentsline{toc}{section}{References}

{%
\scriptsize%
\bibliography{creno}}


\onecolumn%
\addcontentsline{toc}{part}{Document history}
\section*{Document history}

\begin{tabular}{|c|c|c|p{3.5in}|}
 \hline
Version &Date &Author &Details of change \\
 \hline\hline
00A          &29 Sep 2021&Bob Briscoe &First draft.\\\hline%
00B          &01 Oct 2021  &Bob Briscoe &Added geometric interpretation and deterministic case.\\\hline%
00C          &03 Aug 2022  &Bob Briscoe &Added shallow buffer case, empirical results over PIE and discussion of the applicability of the synchronized loss model. Added Acks section. Altered analysis to use max window not min.\\\hline%
00D          &08 Aug 2022  &Bob Briscoe &Added explicit step at \autoref{eqn:p_ratio1} that was previously implicit.\\\hline%
00E          &16 Mar 2023  &Bob Briscoe &Added text around results in \S\,\ref{empirical-pfifo}. Improvements throughout. Expanded conclusions.\\\hline%
\metaversion &\metadate  &Bob Briscoe &Issued without further changes.\\\hline%
\hline%
\end{tabular}

\end{document}